\documentclass[sigconf,screen]{acmart}

\usepackage[utf8]{inputenc}
\usepackage{graphicx}
\newsavebox{\findingboxreg}
\usepackage{algorithm}
\usepackage{algpseudocode}
\usepackage{subcaption}
\usepackage[skip=0pt]{caption}
\usepackage{amsmath}
\usepackage{soul}
\usepackage{xspace}
\usepackage{xcolor}
\usepackage{color, colortbl}
\usepackage{minted}
\usepackage{multirow}
\usepackage[many]{tcolorbox}
\usepackage{enumitem}
\usepackage{ulem}
\usepackage{rotating}
\usepackage[utf8]{inputenc}
\usepackage{cleveref}
\usepackage{framed}
\usepackage{booktabs}
\usepackage{enumitem}
\usepackage{tabularx}
\usepackage{multirow}
\usepackage{wrapfig}
\usepackage{longtable}
\usepackage{listings}

\newcommand{\ttsmall}[1]{\texttt{\small #1}}
\newcommand{\change}[1]{{#1}}
\newcommand{\beaver}{\textrm{Tangent}\xspace}
\newcounter{findingcounter}
\definecolor{findingbg}{RGB}{238,245,252}   
\definecolor{findingaccent}{RGB}{15,98,254} 

\newenvironment{findingbox}[1]{%
  \refstepcounter{findingcounter}%
  \begingroup
  \setlength{\parindent}{0pt}%
  \setlength{\parskip}{2pt}%
  \setlength{\FrameSep}{6pt}%
  \definecolor{shadecolor}{named}{findingbg}%
  \begin{snugshade}%
  {\color{findingaccent}}%
  \textbf{\color{findingaccent}Finding~\thefindingcounter:} \emph{#1.}%
}{%
  \end{snugshade}%
  \endgroup
  \vspace{4pt}%
}

\newenvironment{findingboxnew}[1]{%
  \refstepcounter{findingcounter}%
  \par\vspace{4pt}\noindent
  \setlength{\fboxsep}{6pt}%
  \begin{lrbox}{\findingboxreg}%
    \begin{minipage}{\dimexpr\linewidth-2\fboxsep\relax}%
      \setlength{\parindent}{0pt}%
      \setlength{\parskip}{2pt}%
      \textbf{\color{findingaccent}Finding~\thefindingcounter:} \emph{#1.}\ %
}{%
    \end{minipage}%
  \end{lrbox}%
  \colorbox{findingbg}{\usebox{\findingboxreg}}%
  \par\vspace{4pt}%
}

\newenvironment{dataset}[1]{%
  \begingroup
  \setlength{\parindent}{0pt}%
  \setlength{\parskip}{2pt}%
  \setlength{\FrameSep}{6pt}%
  \definecolor{shadecolor}{named}{findingbg}%
  \begin{snugshade}%
  {\color{findingaccent}}%
  \textbf{\color{findingaccent}\emph{#1.}}%
}{%
  \end{snugshade}%
  \endgroup
  \vspace{4pt}%
}

\lstnewenvironment{pythoncode}
{
  \lstset{
    language=Python,
    frame=lines,
    basicstyle=\ttfamily\scriptsize,
    breaklines=true,
    breakatwhitespace=false,
    numbers=none,
    numbersep=4pt,
    aboveskip=6pt,
    belowskip=6pt,
    columns=fullflexible,
    keepspaces=true,
    showstringspaces=false
  }
}
{}

\newcommand*\Reactivatenumber{%
  \lst@AddToHook{OnNewLine}{%
   \let\thelstnumber\origthelstnumber%
   \advance\c@lstnumber\@ne\relax}%
}
\AtBeginDocument{%
  }

\copyrightyear{2026}
\acmYear{2026}
\setcopyright{cc}
\setcctype{by-nc-nd}
\acmConference[ASE '26]{Proceedings of the 41st IEEE/ACM International Conference on Automated Software Engineering}{October 12--16, 2026}{Munich, Germany}
\acmBooktitle{Proceedings of the 41st IEEE/ACM International Conference on Automated Software Engineering (ASE '26), October 12--16, 2026, Munich, Germany}
\acmDOI{10.1145/3832783.3837414}
\acmISBN{979-8-4007-2882-2/2026/10}

\begin{document}

\title{Tangent: An Empirical Study of Testing Practices for LLM-Based Agent Applications}

\author{Rangeet Pan}
\email{rangeet.pan@ibm.com}
\affiliation{%
  \institution{IBM T.J. Watson Research Center}
  \city{Yorktown Heights}
  \state{NY}
  \country{USA}
}

\author{Tyler Stennett}
\email{tyler.stennett@gatech.edu}
\affiliation{%
  \institution{Georgia Institute of Technology}
  \city{Atlanta}
  \state{GA}
  \country{USA}
}

\author{Divya Sankar}
\email{divya.sankar@ibm.com}

\affiliation{%
  \institution{IBM T.J. Watson Research Center}
  \city{Yorktown Heights}
  \state{NY}
  \country{USA}
}

\author{Bridget McGinn}
\email{bridget.mcginn@ibm.com}

\affiliation{%
  \institution{IBM T.J. Watson Research Center}
  \city{Yorktown Heights}
  \state{NY}
  \country{USA}
}

\author{Alessandro Orso}
\email{orso@uga.edu}
\affiliation{%
  \institution{University of Georgia}
  \city{Athens}
  \state{GA}
  \country{USA}
}
\author{Raju Pavuluri}
\email{pavuluri@us.ibm.com}

\affiliation{%
  \institution{IBM T.J. Watson Research Center}
  \city{Yorktown Heights}
  \state{NY}
  \country{USA}
}
\author{Saurabh Sinha}
\email{sinhas@us.ibm.com}
\affiliation{%
  \institution{IBM T.J. Watson Research Center}
  \city{Yorktown Heights}
  \state{NY}
  \country{USA}
}

\author{Maja Vukovic}
\email{maja@us.ibm.com}
\affiliation{%
  \institution{IBM T.J. Watson Research Center}
  \city{Yorktown Heights}
  \state{NY}
  \country{USA}
}

\renewcommand{\shortauthors}{Pan et al.}

\begin{abstract}

Agents built on large language models (LLMs) are increasingly used to develop applications that perform complex, multi-step tasks involving reasoning, tool use, and interaction with external environments. Despite rapid advances in benchmarking LLM-based agents, little is known about how they are tested in practice. In particular, key aspects of testing---such as testing levels, objectives, data patterns, test complexity, and validation strategies---for agent applications remain underexplored.
In this paper, we present an empirical study of testing practices for LLM-based agent applications using a corpus of mined open-source projects.
We construct an extensive dataset of agent applications, tools, and tests, and manually label 2,572 test methods from 240 modules. From this analysis, we develop a taxonomy of 23 testing patterns across test fixtures, data, objectives, and assertions, and characterize tests by level (unit, module, integration). We complement this with structured interviews of 10 senior industry practitioners building agentic systems.
Our results show that testing of LLM-based agent applications is dominated by narrowly scoped unit tests, with limited coverage of complex interactions, realistic scenarios, and non-functional requirements. Tests often rely on simplistic inputs, heavy mocking, and shallow validation, and exhibit low structural complexity. Although industry practice places greater emphasis on non-functional testing than open-source projects, both reveal common gaps, including the lack of formal testing foundations, unclear test objectives, and challenges in generating high-quality test data.
Based on these findings, we identify research directions toward more systematic and rigorous testing of agent applications, including foundations for agent testability, formalized test objectives, and fault-based testing techniques tailored to agentic systems. We also release a curated dataset of testing artifacts to support future research.
\end{abstract}




\maketitle

\vspace{-5pt}
\section{Introduction}
\label{sec:intro}

Recent advances in large language models (LLMs) have driven a shift from conventional language modeling systems, which statically respond to prompts through single-step inference, toward agentic architectures capable of autonomous or semi-autonomous planning, tool use, and multi-step reasoning~\cite{wang2024survey, xi2025rise}. Through techniques such as chain-of-thought reasoning~\cite{wei2022chain} and the integration of external tools, LLMs can serve as the central component of these systems, coordinating perception, planning, action, and reflection. These capabilities have enabled the emergence of LLM-based agents that perform complex tasks such as software engineering, web navigation, and autonomous exploration (e.g., ~\cite{liu2024large, deng2023mind2web, zhou2023webarena}).

The growing autonomy and complexity of LLM-based agents make rigorous testing increasingly critical. However, this is challenging due to the wide range of complex behaviors that agents can exhibit over long interactions. They must be assessed across multiple dimensions, including planning, tool use, memory management, reflection, adaptation, environment interaction, robustness, and safety. Multi-agent systems introduce an additional level of complexity through inter-agent interactions that must also be validated. 
Addressing these challenges requires suitable validation techniques that can be applied across these dimensions and at different testing levels (e.g., unit, integration, and end-to-end), along with meaningful test adequacy metrics. 
Without sufficient testing, failures can range from minor incidents (e.g., unintended giveaways~\cite{wsjplaystation}) to serious security breaches involving sensitive data~\cite{metasecuritybreach}.

Existing work in this area has largely focused on developing benchmarks for evaluating LLM-based agents across a range of tasks with varying levels of complexity (e.g.,~\cite{jimenez2023swebench, aleithan2024swebench_plus, yang2026swesmith, deng2025swebench_pro}) and analyzing agent trajectories  (e.g.~\cite{kim2025beyond, bouzenia2025understanding, qian2025webgrapheval}). However, comparatively less attention has been paid to systematic testing and validation approaches for LLM-based agents, particularly techniques for analyzing agent behavior and identifying failures, with only a limited number of such methods proposed to date~\cite{kohl2025automated, ma2025rethinking} and simple tool support for test generation~\cite{agenttoolkit}.
Developing such techniques is essential for improving the reliability, robustness, and trustworthiness of LLM-based agent systems. Moreover, these techniques should be grounded in, and informed by, developer testing practices, which can provide practical insights into how such systems are built and evaluated in real-world settings.

In this work, we present an empirical study of developer testing practices for LLM-based agent applications, with the goal of understanding how developers validate agent behavior in practice. We begin by mining open-source GitHub repositories to construct a dataset of LLM-based agent applications. Our study focuses on Python applications built using popular agent-development frameworks, including LangChain~\cite{langchain}, LangGraph~\cite{langgraph}, and MCP~\cite{mcp}. Using a set of selection criteria, we identified 1,190 candidate repositories, which were then processed through an extraction and filtering pipeline to identify agent-related test modules and generate multiple dataset snapshots. In the final step, the extracted test modules were manually reviewed and labeled, resulting in a dataset of 2,572 test methods annotated along different dimensions (\S\ref{sec:dataset}). To support reproducibility and future research on testing of agent applications, we make the dataset publicly available~\cite{tangent_artifact}.

Using these dataset snapshots, we investigate four research questions aimed at characterizing how developers design tests for agent systems and the techniques and patterns they employ. The first characterizes the testing levels (e.g., unit, integration) and test types (functional or non-functional) used in practice (\S\ref{sec:rq_agent_test_types}). Testing is predominantly unit-level: most tests (61.8\%) exercise individual tools or agents in isolation, about a third target interaction behavior, and only 7.5\% address non-functional requirements such as security and performance. Testing practices also vary with the agent-development framework and align with its level of abstraction. The second examines testing patterns (\S\ref{sec:rq_agent_test_patterns}), for which we develop a catalog of 23 patterns spanning test fixtures, data, objectives, and assertions. The results expose narrow scoping throughout: mocking is widespread, 39.9\% of tests rely on simple example-like inputs, and few tests exercise complex scenarios, execution environments, or system boundaries. The third analyzes the structural complexity of agent-related tests, which we find to be low (\S\ref{sec:rq_agent_test_complexity}). Lastly, the fourth draws on structured interviews with industry practitioners (\S\ref{sec:rq_industry}), which corroborate many of our findings and surface challenges not observable from test artifacts alone.

Leveraging the insights gained from this study, we outline several research directions to advance systematic and rigorous testing of agent applications (\S\ref{subsec:test-level-future-direction}, \S\ref{subsec:test-pattern-future-direction}). The first direction is the development of new testability theory and foundations that extend notions of controllability and observability to agent reasoning, tool usage, non-determinism, adaptive behavior, and environmental interactions. Another direction involves formalization of testing objectives for agent-based systems, developing notions of test adequacy at different testing levels that account for agentic abstractions such as tool-selection paths, reasoning traces, delegation patterns, and multi-step task execution. A further direction is the development of fault-based testing techniques tailored to agentic systems, including agent-specific fault models
and corresponding techniques for fault injection and detection. Additional directions include approaches for improving test isolation, enabling more comprehensive validation, and systematically generating test inputs and environments.


\change{Our work complements \citeauthor{hasan2025empirical}~\cite{hasan2025empirical}, who catalog setup and verification patterns across agent frameworks and applications and map them to canonical architectural components. We instead characterize test design itself---testing levels and types, objectives, data, isolation boundaries, and structural complexity---at a finer granularity, and extend repository evidence with practitioner interviews and a research roadmap. We detail this comparison in \S\ref{sec:related_works}.}

The key contributions of this work include:

\begin{itemize}[leftmargin=*, topsep=2pt]

\item A systematic study of testing practices for LLM-based agent applications using a large corpus of open-source projects and industry practitioner interviews.

\item A comprehensive analysis and discussion of common testing strategies and patterns used in practice, the key limitations of current approaches, and research directions for advancing the testing and validation of agent applications.

\item An artifact consisting of a dataset of labeled agent-related test cases and analysis tools to support replication of our work and future research on agent testing and validation~\cite{tangent_artifact}.

\end{itemize}





\section{Dataset Collection and Labeling}
\label{sec:dataset}

\begin{figure*}[t]
    \centering
    \includegraphics[width=.74\linewidth]{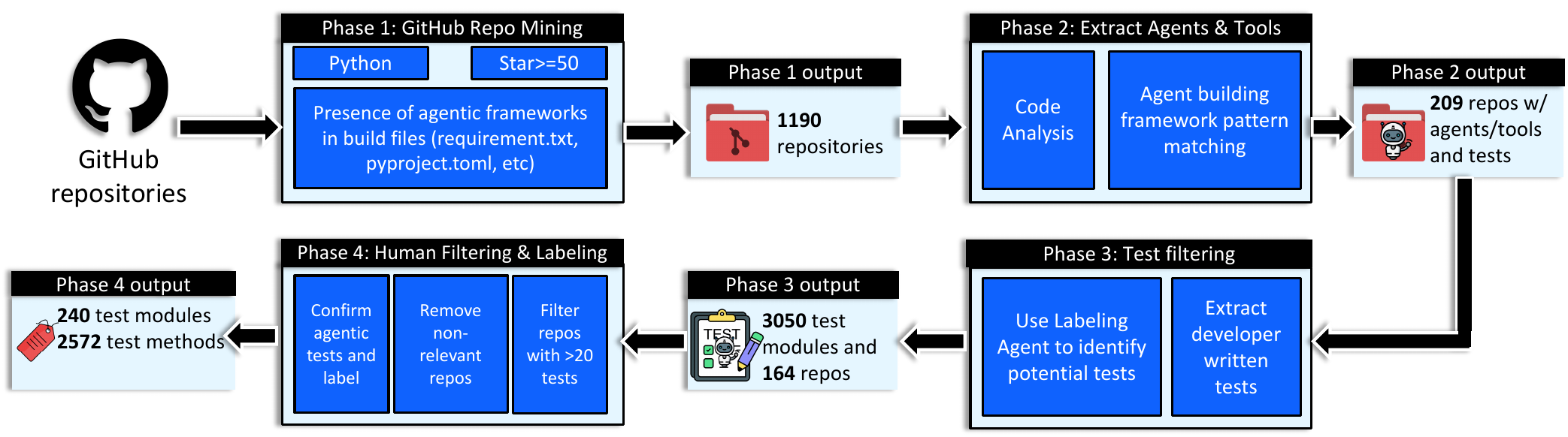}
    \caption{The \beaver dataset collection and test labeling steps.}
    \label{fig:dataset}
    \vspace{-12pt}
\end{figure*}

In this section, we discuss the data collection process and the method used to
identify and confirm agent- or tool-related tests and extract their characteristics. As shown in Figure~\ref{fig:dataset}, the approach has four steps: (1) mining GitHub repositories to build a corpus of candidate projects, (2) using static analysis to detect projects with agents and retain those with tests, (3) filtering the detected tests for relevance, and (4) manually verifying and labeling the selected tests across multiple dimensions.

\vspace{-6pt}
\paragraph{Phase 1: Repository Mining}
We began by mining GitHub repositories whose primary programming language is Python.
\change{This scope reflects the composition of the agentic ecosystem, which is predominantly Python-based~\cite{wang2025empirical}.}
To improve project quality, we applied additional filtering criteria, including a minimum star count~\cite{borges2018s} of 50. We next examined build files (e.g., \ttsmall{requirements.txt}, \ttsmall{setup.py}, and \ttsmall{pyproject.toml}) to identify projects that depend on commonly used agent or tool-building frameworks. 
To avoid GitHub search-result caps, we split the search into quarterly time windows beginning on January 1, 2024\change{, and ending on February 16, 2026, when the dataset was initially collected}. 
This process yielded 1,190 repositories using 11 agent or tool-building frameworks.

\vspace{-6pt}
\paragraph{Phase 2: Agent and Tool Extraction}
Next, we performed static analysis to identify code segments that implement agents or tools. We used Scalpel~\cite{li2022scalpel, scalpel} to extract symbol table information from each codebase, which we used to populate the analysis code model detailed in our artifact~\cite{tangent_artifact}.
To detect agent and tool implementations, we reviewed each framework's documentation to identify its characteristic usage patterns. \change{Although the frameworks differ in their APIs, they often share a common structure: agents are instantiated through framework-specific constructors, tools are defined as decorated or plain functions or classes, and tools are attached to agents through explicit binding mechanisms.} We encoded these patterns as detection rules over framework-specific imports and call-site information, and matched them against the extracted code model to determine whether a codebase creates or uses agents or tools.
The full set of patterns is available in our artifact~\cite{tangent_artifact}. Using them, we identified 1,723 agents, 4,111 tools, and 324 repositories containing agent usage, tool usage, or both.
Among the repositories containing agents or tools, 209 include test cases. 

\vspace{-5pt}
\paragraph{Phase 3: Test Identification and Filtering}
Within the 209 repositories, we identified 17k+ test modules. As manually reviewing all modules was infeasible, to reduce this effort and focus on the most relevant tests, we employed an agentic labeling approach consisting of two labeling agents and one adjudicator agent. The labeling agents determined whether a test module contained indicators that it tests agents or tools. The prompts instructed the agents to be permissive to minimize false negatives, at the cost of some false positives, encouraging them to flag potentially relevant tests. When the two labeling agents produced conflicting labels, the adjudicator agent resolved the conflict. We implemented the agents using Gemini models~\cite{gemini_models}---Gemini 2.5 Flash for the labeling agents and Gemini 2.5 Pro for the adjudicator. The implementation is available in our artifact~\cite{tangent_artifact}. This process identified 3,050 relevant test modules from 164 repositories, which were then manually examined to ensure that they are indeed related to agents and/or tools.
\vspace{-5pt}
\paragraph{Phase 4: Manual Filtering, Verification, and Labeling}
To focus the manual analysis on repositories with substantial testing activity and comprehensive agent-related test suites, we restricted the dataset to applications containing at least 20 test modules. We also excluded repositories that primarily served as agent-building examples, frameworks, \change{or projects focused on tasks peripheral to agent functionality, such as agent sandbox creation and other supporting infrastructure}. These criteria yielded 19 projects with 1,261 test modules, which one labeler manually reviewed to identify those related to agents, tools, or both, resulting in 240 relevant test modules from 12 projects. In the formal labeling phase, two authors independently labeled each test method, verified that it was agent-related, and met regularly to resolve disagreements label by label. Cohen's kappa coefficient~\cite{cohen1960coefficient} was 0.92, indicating near-perfect agreement.
At the end of this phase, we retained 240 test modules with 2,572 test methods that test LLM-based agents or tools. \change{Table~\ref{tb:dataset} lists each project's module and method counts, application domain, and agent-building library versions.
Five projects embed agents within end-user solutions (AI Investment Agent, AIAgents4Pharma, Docmind-AI-LLM, OpenChatBI, and RF-MCP), three provide orchestration runtimes (Agentic Fleet, AgentPool, and Solace Agent Mesh), two support deployable agents (Agentic Fleet and Upsonic), two support deployable tools (Any Agent and Maverick-MCP), and Intelli focuses on agent-workflow creation.
In this study, we use the term ``LLM-based agent application'' broadly to include systems that provide reusable agents or tools, embed agents in domain-specific solutions, or orchestrate agent-based workflows.}

\begin{table}[t]
\centering
\caption{\change{Dataset used for manual labeling.}}
\label{tb:dataset}

\scriptsize
\setlength{\tabcolsep}{2.5pt}
\renewcommand{\arraystretch}{1.12}

\begin{tabularx}{0.9\columnwidth}{@{}
>{\raggedright\arraybackslash}p{1.75cm}
>{\centering\arraybackslash}p{0.55cm}
>{\centering\arraybackslash}p{0.58cm}
>{\raggedright\arraybackslash}p{1.15cm}
>{\raggedright\arraybackslash}X
@{}}
\toprule
\textbf{Proj.} &
\textbf{Mod.} &
\textbf{Met.} &
\textbf{Domain} &
\textbf{Library Version(s)} \\
\midrule

Agentic Fleet~\cite{agentic-fleet}
& 7 & 63
& Generic
& MCP 1.25.0; DSPy 3.1.0b1; OpenAI Agents 0.6.4 \\

AgentPool~\cite{agentpool}
& 30 & 183
& Generic
& MCP 1.26.0; FastMCP 3.1.0 \\

AI Investment Agent~\cite{ai-investment-agent}
& 15 & 300
& Finance
& LangChain 1.2.18; LangGraph 1.1.10 \\

AIAgents4Pharma~\cite{aiagents4pharma}
& 29 & 77
& Pharma
& LangChain 0.3.7; LangGraph 0.3.34 \\

Any Agent~\cite{any-agent}
& 19 & 63
& Generic
& LangChain 1.3.11; LangGraph 1.2.7; MCP 1.28.1;
  LlamaIndex 0.14.23; OpenAI Agents 0.17.7 \\

Docmind-AI-LLM~\cite{docmind-ai-llm}
& 11 & 107
& Document Processing
& LangChain 1.2.6; LangGraph 1.0.6; LlamaIndex 0.14.12 \\

Intelli~\cite{intelli}
& 19 & 48
& Generic
& MCP 1.9.4 \\

Maverick-MCP~\cite{maverick-mcp}
& 21 & 304
& Finance
& LangChain 1.1.3; LangGraph 1.0.4; MCP 1.15.0;
  FastMCP 2.10.6 \\

OpenChatBI~\cite{openchatbi}
& 8 & 103
& Business Intelligence
& LangChain 0.3.27; LangGraph 0.6.7; MCP 1.13.1 \\

RF-MCP~\cite{rf-mcp}
& 14 & 173
& Test Automation
& MCP 1.22.0; FastMCP 2.13.3 \\

Solace Agent Mesh~\cite{solace-agent-mesh}
& 19 & 381
& Generic
& MCP 1.24.0 \\

Upsonic~\cite{upsonic}
& 48 & 770
& Generic
& MCP 1.26.0; FastMCP 2.14.5 \\

\midrule
\textbf{Total}
& \textbf{240}
& \textbf{2572}
& --
& -- \\
\bottomrule
\end{tabularx}

\vspace{2pt}
\raggedright
\textit{Proj.:} Project name;
\textit{Mod.:} number of modules;
\textit{Met.:} number of methods.
\end{table}

\vspace{-3pt}
\begin{dataset}{\beaver Dataset}
This work contributes a dataset containing 1,723 agents and 4,111 tools, along with 240 manually labeled test modules consisting of 2,572 agent-related test cases.
\end{dataset}
\vspace{-7pt}

\vspace{-7pt}
\section{Agent-Specific Testing Levels and Test Types}
\label{sec:rq_agent_test_types}

\begin{description}[leftmargin=1.0em, labelsep=0.5em, itemsep=0pt, topsep=2pt]
    \item[RQ1 (Testing Level):] How do developers test agent applications in practice in terms of testing levels and test types?
\end{description}

In conventional software testing, levels such as unit and integration are well defined. Agentic systems add new requirements, as components must be tested both in isolation and in interaction with other agents and non-agentic components.
\change{In this work, we define testing levels according to the granularity of the focal test target, consistent with conventional software testing~\cite{ammann2016testing,iso29119}. Thus, the testing level is determined by what is exercised, not by how.}
This section presents the testing levels for agentic systems, summarizes our observations, and identifies future research directions.


\vspace{-5pt}
\subsection{Definitions and Illustrative Examples}

\paragraph{Unit Testing}
\change{
We define unit testing for agentic systems as evaluating the smallest independently testable element. A unit may be an individual tool, an agent, or an internal subcomponent, such as memory, planning, or reasoning. Although unit tests are often executed in isolation using test doubles (e.g., mocks and stubs)~\cite{mackinnon2000endo, meszaros2007xunit}, in this work, we treat isolation as an orthogonal test-design decision rather than a defining characteristic of unit testing. 
}

For example, in MCP-based applications~\cite{mcp}, tools are commonly deployed as services on an MCP server that LLMs connect to for invocation. The following example from Maverick MCP~\cite{maverick-mcp} illustrates unit testing of a tool: the \ttsmall{data\_fetch\_stock\_data} tool is executed with sample data, and its output is checked for correctness. Before this validation, the test performs setup steps such as creating a test database or mocking external services, similar to the setup phase in conventional software testing. In this case, the focal unit under test is the tool.

    \begin{pythoncode}
async def test_fetch_stock_data(self, test_db, mock_redis):
"""Test fetching stock data from the database."""
    async with Client(mcp) as client:
        result = await client.call_tool("/data_fetch_stock_data",{"request": {"ticker": "AAPL","start_date": "2024-01-01","end_date": "2024-01-31",}},)
        assert len(result) > 0...
    \end{pythoncode}


The focal unit may also be an agent rather than a tool. For example, the following test from the AgentPool project~\cite{agentpool} asserts that the \ttsmall{AGUIAgent} can correctly register tools, treating the agent as an indivisible test target.


\begin{pythoncode}
async def test_agui_agent_register_tool():
    def my_tool(text: str) -> str:
        return text.upper()
    async with AGUIAgent(endpoint="http://localhost:8000/run", name="test-agent") as agent:
        tool = agent.tools.register_tool(my_tool)
        assert tool.name == "my_tool"...
\end{pythoncode}

\vspace{-5pt}
\paragraph{Module Testing}

\change{
In conventional software testing, a module test validates a software module composed of multiple interacting units, checking both its externally observable behavior and the interactions among its units. For example, a module test may invoke a sequence of methods to drive an object through a series of state transitions, verifying its behavior at each stage. An analogous abstraction applies to agentic systems, where the focal target is typically an entire agent. Unlike unit testing, which evaluates the smallest independently testable element, module testing treats the agent as a composition of interacting capabilities—such as reasoning, planning, tool use, and memory. An agent may therefore serve as the focal target at either testing level: as a unit when treated as an indivisible test target (typically for simple agent implementations), or as a module when the objective is to validate the interactions among its internal capabilities. Thus, a module test may drive an agent through a sequence of execution states by assigning tasks and controlling the available tools and execution environment, enabling systematic validation of the agent's behavior, its state transitions, and the coordination among its internal capabilities.
}

The following test from Upsonic~\cite{upsonic} evaluates an agent in a synthetic scenario where two tools are registered and invoked. The test validates that the agent behaves correctly under multi-tool interaction. Such tests are important for uncovering defects that arise from tool composition---particularly issues in tool sequencing and interaction---and their effects on the agent’s state, including its outputs, intermediate reasoning traces, and memory.


    \begin{pythoncode}
async def test_e2e_multiple_task_tools_execution():
    agent = Agent(...)
    task = Task( description="Use add_numbers to calculate 10 + 5, and then use multiply_numbers to calculate 3 * 7. Return both results separated by comma.", tools=[add_numbers, multiply_numbers])
    ...
    response_text = str(task.response) if task.response else ""
    assert "15" in response_text, f"Should contain '15' (from 10+5), got: ..."
    assert "21" in response_text, f"Should contain '21' (from 3*7), got: ..."
    \end{pythoncode}
    
\vspace{-5pt}
\paragraph{Integration Testing}

In conventional software systems, integration testing evaluates whether interacting components work together as expected, using either mocks or real components. In agentic systems, the concept is similar but often involves more complex interactions, including multiple agents, agents interacting with non-agentic components, or both. For example, a conventional component may provide input to an agent, an agent’s output may be consumed by a non-agentic component, or multiple agents may collaborate by exchanging intermediate results to produce a final outcome. The following test case from Intelli~\cite{intelli} illustrates this complexity: a travel-assistant workflow orchestrates nine agents across text, speech, image generation, and vision to generate a complete travel itinerary. The test validates several outputs, including the itinerary, synthesized audio, generated images, and the final travel guide. This example highlights how agentic systems combine heterogeneous components through dynamic workflows.

    \begin{pythoncode}
def test_travel_assistant_multimodal_flow(self): ...
    # 1. Travel Itinerary Creation
    itinerary_agent = Agent(mission="Create a detailed 3-day travel itinerary",)
    itinerary_task = Task(TextTaskInput("Create a 3-day travel itinerary for Rome, Italy. Include major attractions, food recommendations, and transportation tips."),...
    # 2. Speech Synthesis
    speech_agent = self._create_speech_agent()
    speech_task = Task(TextTaskInput("Convert the first day of this itinerary to speech for the traveler"),..)...
    # 9. Final Travel Package
    final_agent = Agent(mission="Create an enhanced travel package combining all insights",)...
    self.assertEqual(results["final_package"]["type"],"text")
    \end{pythoncode}
    
\vspace{-5pt}
\paragraph{API Testing}

Similar to conventional systems, agentic systems are often deployed as services and validated via their API endpoints. In this setting, testing focuses on checking whether the system correctly exposes and orchestrates its agentic components through the API. For example, the following test from the AgenticFleet application~\cite{agentic-fleet} validates that agents are properly registered and accessible via the system’s endpoint. 

    \begin{pythoncode}
def test_get_agents(client: TestClient, mock_workflow: MagicMock):
    agent1 = MagicMock()
    agent1.name = "agent1" ...
    agent2 = MagicMock()
    ...
    mock_workflow.agents = {"agent1": agent1, "agent2": agent2}
    response = client.get("/api/v1/agents")
    assert response.status_code == 200
    \end{pythoncode}

\vspace{-5pt}
\paragraph{End-to-end Testing}
Tests that exercise flows across all layers of an agent-based application---from user input to final system output---can be considered end-to-end. Some tests we labeled as integration tests may fall into this category, but we could not confirm this from repository analysis alone, as we did not deploy the applications  to verify whether the tests exercise end-to-end-flows. Therefore, we explicitly asked industry practitioners about end-to-end testing, as they work with deployed agentic systems and are better positioned to characterize this testing level.

\begin{figure*}[t]
    \centering
    \includegraphics[width=0.7\linewidth]{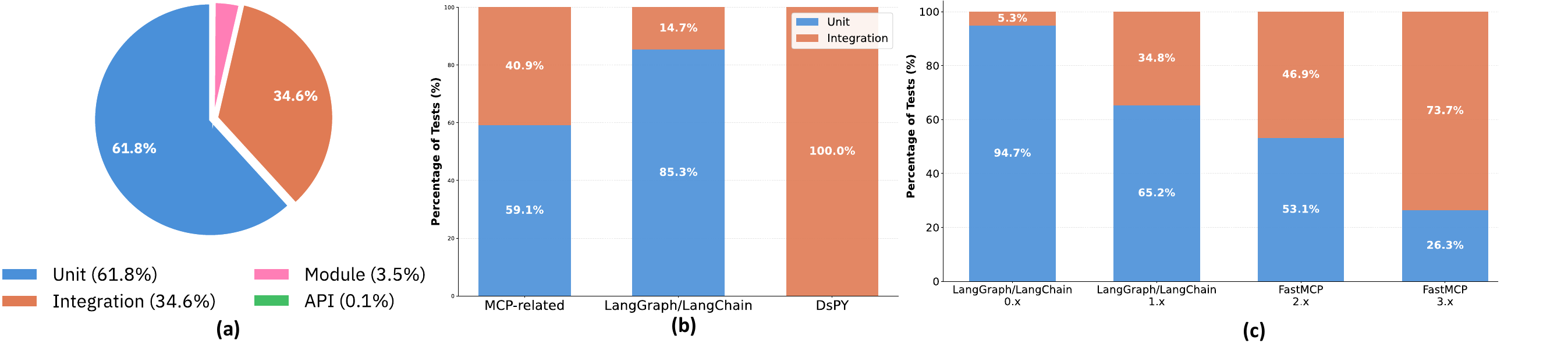}
    \vspace{-3pt}
    \caption{Distribution of test types: (a) overall, (b) across different frameworks, (c) different versions of frameworks.}
    \label{fig:test_level_pattern}
    \vspace{-11pt}
\end{figure*}

\vspace{-6pt}
\subsection{Key Observations and Findings}

Figure~\ref{fig:test_level_pattern}(a) shows the distribution of test types in the manually labeled dataset. Most tests (61.8\%) evaluate tools or agents in isolation, while 34.6\% target interactions among multiple agents or between agents and other components. Only a small fraction cover multi-tool invocation (3.5\%) or application behavior through exposed APIs (0.1\%). A closer analysis reveals several patterns, including framework effects on testing levels, project-specific testing behavior, and a notable lack of tests for non-functional requirements.

\vspace{-5pt}
\begin{findingbox}{Testing Level}
Most tests (61.8\%) focus on tools or agents in isolation, with fewer tests (34.6\%) targeting interaction-level behavior; only a small fraction evaluates multi-tool scenarios (3.5\%) or assesses agentic systems via exposed API endpoints (0.1\%).
\end{findingbox}

\vspace{-13pt}
\paragraph{Effect of Agent-development Framework on Testing Levels}
\label{par:agent_framework_on_test_level}
To examine this effect, we compared the frameworks in the manually labeled dataset with the distribution of test types, excluding module- and API-level tests because of their small numbers. Figure~\ref{fig:test_level_pattern}(b) shows the results. We grouped related frameworks, such as MCP with FastMCP and LangGraph with LangChain, because they are often used together. 
The results suggest that frameworks lie along a spectrum of programming abstraction, and testing practices are significantly associated with that spectrum (p < 0.001).
At one end, MCP~\cite{mcp} and FastMCP~\cite{fastmcp} provide higher-level abstractions with clear isolation boundaries around tools and agents. As a result, applications built with these frameworks favor unit testing, since developers can validate well-defined components without instantiating the full system. However, this also shifts some integration-testing burden to end users who compose tools through MCP servers.
In the middle, LangGraph~\cite{langgraph} and LangChain~\cite{langchain} require more explicit definitions of tools, agents, and coordination. Applications built on these frameworks therefore rely more on integration testing, as correctness depends on both component behavior and their interactions.
At the other end, DSPy~\cite{dspy} adopts a lower-level model in which developers are responsible for composition and orchestration. Because boundaries are less clear, isolation is harder. Consequently, these applications rely even more heavily on integration testing.


\change{
We next examine whether this pattern also holds across the major library versions used by the projects in our dataset. Figure~\ref{fig:test_level_pattern}~(c) shows the distribution of unit and integration tests for the two major versions of LangChain, LangGraph, and FastMCP present in our dataset. For each library, projects using the newer major version consistently exhibit a higher proportion of integration tests.
A plausible explanation is the evolution of programming abstractions in agent-building frameworks. For example, LangChain~0.x exposes multiple abstractions for building LLM applications and agents, including chains, agent executors, manually bound tools, prompt objects, and LangGraph prebuilt agents.
In contrast, LangChain~1.x consolidates these concepts into a smaller, agent-centered API. These changes shift client applications from manually composing chains, executors, hooks, and tool-bound models to using a framework-managed agent loop with explicit tools, typed state, and structured output.
As a result of these abstraction changes, developers may increasingly rely on integration tests to validate end-to-end agent behavior instead of unit tests for individual components.
}

\vspace{-5pt}
\begin{findingboxnew}{Agentic Framework vs. Testing Level}
The distribution of testing levels varies by framework and appears to correlate with framework abstraction. Applications built with MCP and FastMCP tend to favor unit testing, applications using LangChain and LangGraph tend to balance unit and integration testing, and applications built with DSPy tend to rely more heavily on integration testing. \change{We also observe a shift from unit testing to integration testing in newer framework versions, consistent with higher abstraction levels.}
\end{findingboxnew}
\vspace{-4pt}


\begin{wrapfigure}{r}{0.44\columnwidth}
    \centering
    \vspace{-8pt}
    \includegraphics[width=\linewidth]{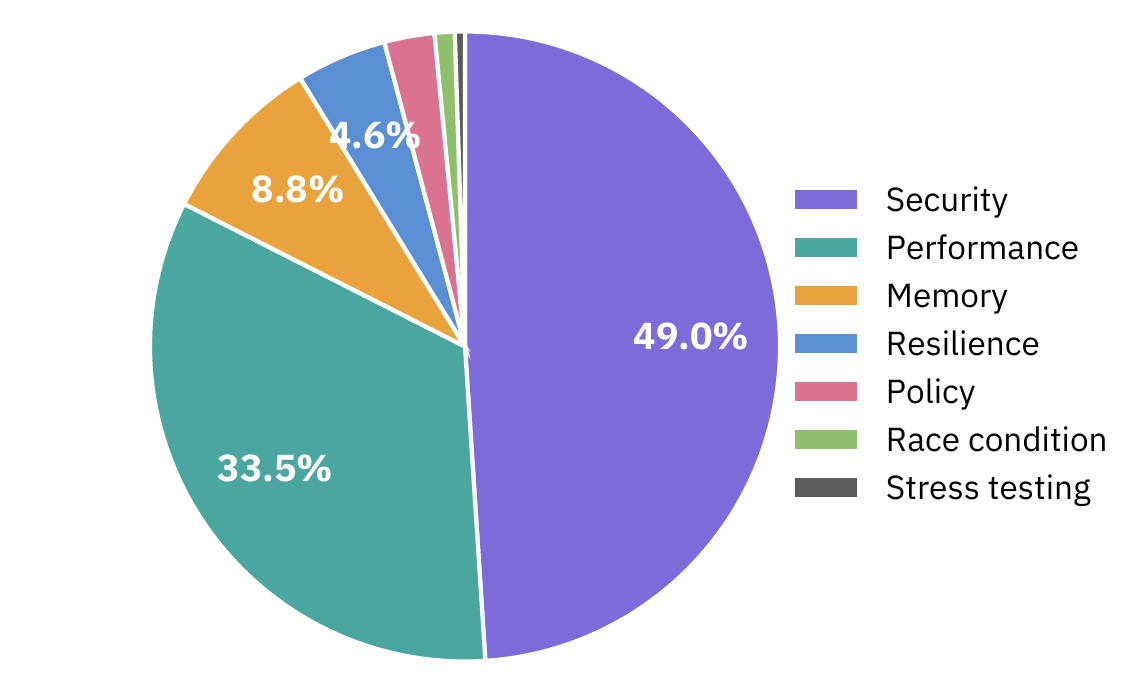}
    \caption{Distribution of NFR testing types.}
    \label{fig:nfr}
    \vspace{-8pt}
\end{wrapfigure}
\vspace{-9pt}
\paragraph{Lack of Non-Functional Testing.}
{\emergencystretch=3em
Only 7.5\% of the manually labeled tests target non-functional requirements (NFRs), indicating that such testing is uncommon in agent applications. As shown in Figure~\ref{fig:nfr}, among the NFR tests, the majority focus on security (49.0\%), followed by performance (33.5\%), memory (8.8\%), and policy-related concerns (4.6\%). NFR testing is highly concentrated: among the 12 projects we examined, Upsonic~\cite{upsonic} alone accounts for 66.5\% of all such tests, primarily targeting security scenarios. For example, the following test verifies that prescription numbers in prompts are anonymized before LLM processing.\par}

\begin{pythoncode}
async def test_medical_info_anonymize_prescription(mock_infer_model):
    mock_response = ModelResponse(parts=[TextPart(content="...")], ...)
    agent_with_anonymize_policy = Agent(user_policy=MedicalAnonymizePolicy,...)
    prescription_task = Task(description="Prescription number RX123456789 ...")
    assert "RX123456789" not in result
\end{pythoncode}
More generally, the scarcity of NFR-focused tests indicates that such concerns remain under-tested, despite their importance in agentic systems with broad access to external tools and data.



\vspace{-3pt}
\begin{findingboxnew}{Non-Functional Testing}
    NFR-focused testing is limited in agentic projects: only 7.5\% of the labeled tests target NFRs and just 33.5\% of these address performance, indicating that performance and other NFRs remain significantly under-tested in practice.
\end{findingboxnew}
\vspace{-9pt}

\subsection{Future Research Directions}
\label{subsec:test-level-future-direction}

Next, we outline research directions to address the gaps seen in how agentic systems are currently tested. Our perspective is grounded in the observation that many of the challenges in agentic systems resemble problems that have long been studied in the SE literature.
\vspace{-5pt}
\paragraph{Formalizing Test Objectives}
Systematic testing depends on a shared vocabulary, including testing levels, adequacy criteria, and quality characteristics, that is well established in conventional standards and glossaries~\cite{iso24765, iso25010:2023, iso29119, isqtb_glossary}. Recent extensions address AI-specific concerns such as explainability and non-determinism~\cite{iso25059, istqb_ct_ai, dobslaw2025challenges}, but they do not yet define testing levels or adequacy criteria for agentic systems, where behavior emerges through the dynamic orchestration of tool use, reasoning, memory, and delegation.
Here, we use conventional testing levels (e.g., unit, module, integration, and API) as an initial lens for studying agentic systems while exposing their limitations. The boundary between functional and non-functional behavior often blurs, conventional hierarchies assume stable module interfaces while agent boundaries are fluid and framework-dependent, and existing adequacy criteria such as statement or branch coverage do not capture agent-specific behaviors like tool selection, or reasoning traces. These observations suggest the need for a new testing ontology for agentic systems, together with adequacy criteria that reflect what is being evaluated, under which conditions, and at which architectural boundary.
\vspace{-5pt}
\paragraph{Testing Non-functional Requirements} 
Our findings show that only a small fraction of tests target NFRs. Conventional SE offers well-established techniques for evaluating behavior under failure---fault injection~\cite{segall1995fiat, arlat1990fault} and chaos engineering~\cite{basiri2016chaos}---and recent work has begun extending them to agents, cataloging behavioral failure modes in multi-agent traces~\cite{cemri2026multi}, injecting infrastructure-level faults such as timeouts and schema drift into single-agent systems~\cite{gupta2026reliabilitybench}, and defining fault taxonomies for reasoning and conversation history in multi-agent settings~\cite{jia2026mas}.
However, these efforts remain partial. To our knowledge, no prior work provides a fault taxonomy or injection mechanism that spans the full agentic stack, including tool integration, memory, inter-agent communication, and security, or that models fault propagation across these layers.
\change{The scarcity of NFR tests may thus reflect not only missing test mechanisms but also the difficulty of defining concrete non-functional targets: properties such as robustness, security, and recoverability can depend on interactions across architectural layers, obscuring what to test and under which failure conditions. Future work should develop layer-aware fault models and injection mechanisms that make such objectives concrete and measurable.}

\vspace{-5pt}
\paragraph{Testability Theory for Agentic Frameworks}
Our finding that testing-level distributions correlate with framework abstraction raises a deeper question: what framework properties make agentic systems more or less testable? Software testability is well studied. Freedman~\cite{freedman1991testability} defines it in terms of observability and controllability, Voas and Miller~\cite{voas2002software} formalize it as the likelihood that a fault causes a failure during testing, and Binder~\cite{binder1994design} translates these ideas into design-for-testability patterns. Although these concepts have been applied across many software domains~\cite{garousi2019survey}, no prior work defines testability for agentic frameworks or relates framework design choices to observed testing practices.
In agentic systems, observability extends beyond program variables to memory state, tool selection, delegation, and intermediate reasoning artifacts. Controllability likewise goes beyond supplying inputs: it may require stubbing model outputs, intercepting tool responses, and injecting failures. The near-absence of module-level testing in our dataset may therefore reflect not only developer preference, but also limited controllability and observability at the level of individual agents. When frameworks do not expose interfaces for isolating, observing, and controlling agent behavior, developers are pushed toward system-level testing.
We therefore argue that testability in agentic frameworks should be defined through dimensions that extend the classical controllability-observability view. These include whether agents or tools can be isolated without instantiating the full system, how much internal execution state is observable, how much control developers have over model outputs, tool responses, and environment conditions, whether intermediate states can be asserted, and whether traces can be replayed with sufficient determinism. 

\vspace{-3pt}
\section{Agent-Specific Testing Patterns}

\label{sec:rq_agent_test_patterns}
\begin{figure*}[!htp]
    \centering
    \includegraphics[width=.7\linewidth]{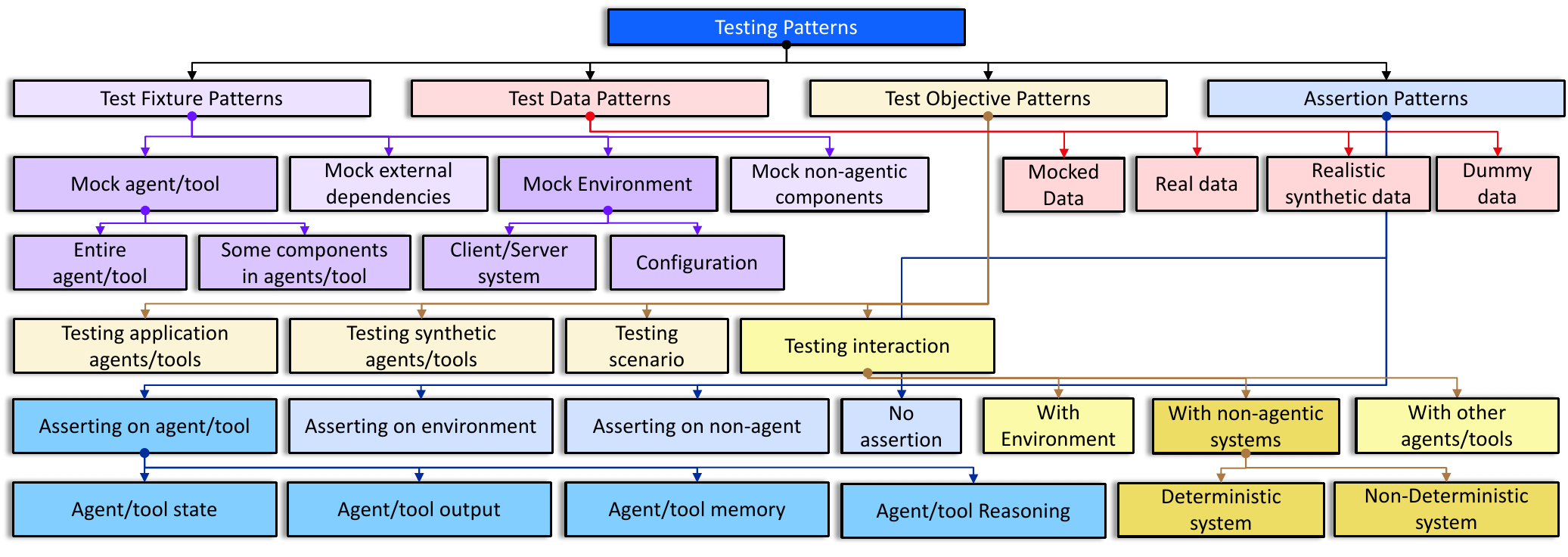}
    \caption{The testing patterns identified from our dataset. }
    \label{fig:test_pattern}
    \vspace{-10pt}
\end{figure*}

\begin{description}[leftmargin=1.0em, labelsep=0.5em, itemsep=0pt, topsep=2pt]
    \item[RQ2 (Testing Patterns):] What are the different testing patterns that developers employ in practice to test agent applications?
\end{description}

To investigate this research question, we developed a classification scheme for the testing practices observed in the manually examined tests. \change{We structured the scheme around four predefined categories that reflect the main constituents of a test}: \textit{test fixture patterns}, \textit{test data patterns}, \textit{test objective patterns}, and \textit{test assertion patterns}. Together, these categories capture how tests isolate components, what inputs they use, what they validate, and how correctness is checked. \change{Within each category, the individual patterns emerged inductively during inspection and were iteratively refined as we examined additional tests.} Overall, we identified 23 patterns; Figure~\ref{fig:test_pattern} shows them, organized into the four categories.

\vspace{-5pt}
\subsection{Definitions and Illustrative Examples}
\label{subsec:test-pattern-definition}

\paragraph{Test Fixture Patterns} In these patterns, we specifically examine which agentic components are mocked. For instance, tests may replace entire agents or tools with mocks to evaluate system behavior under controlled conditions. In the following example from the AI Investment Agent project~\cite{ai-investment-agent}, the \ttsmall{fetch\_reference\_content} tool is mocked, and the pipeline’s output is validated accordingly.

    \begin{pythoncode}
async def test_execute_tool_calls_caps_per_turn(self):
    mock_tool = AsyncMock()...
    editor._tools_by_name = {"fetch_reference_content": mock_tool}
    tool_calls = [{"name": "fetch_reference_content", "args": {"url": f"https://example.com/{i}"}, "id": f"call_{i}",} for i in range(5)]
    results = await editor._execute_tool_calls(tool_calls)
    assert len(results) == 5
    \end{pythoncode}
    
In other cases, specific parts of an agent (e.g., LLM call, reasoning, memory) may be mocked. In the following example from the Maverick MCP application~\cite{maverick-mcp}, the test mocks only the query-classification functionality rather than the entire supervisor agent.
    
    \begin{pythoncode}  
async def test_multi_agent_parallel_routing(self, supervisor_agent):
    # Mock classification for investment decision
    supervisor_agent.query_classifier.classify_query = AsyncMock(return_value={"category": "stock_investment_decision", ...})
    result = await supervisor_agent.coordinate_agents(query="Should I buy AAPL for my moderate risk portfolio?",)
    assert result["status"] == "success"
    \end{pythoncode}

\vspace{-5pt}
\paragraph{Test Data Patterns}
Test data is central to testing, and this pattern category examines the kinds of inputs tests use. These may include mocked objects, real data, or synthetic but realistic data representing typical scenarios. We also found a pattern in which tests use simple placeholder inputs \change{(or ``dummy data'')} such as \ttsmall{test1}, \ttsmall{test2}, etc. Although such inputs may be sufficient for basic use cases, they generally do not represent real-world usage and thus provide limited support for more rigorous testing.


\vspace{-5pt}
\paragraph{Test Objective Patterns}
\label{par:test_obj_patterns}
\change{Next, we examine test validation objectives along three independent dimensions: the subject under test, how it is exercised, and the interactions validated. The subject may be an application-defined or synthetic agent or tool. It may be exercised through a single invocation or a multi-step scenario. Tests may also validate interactions with deployment environments, other agents or tools, or non-agentic components. These dimensions are independent and their patterns may overlap.
} 



\vspace{-5pt}
\paragraph{Test Assertion Patterns}
Finally, we examine how assertions are expressed in tests. Some tests assert directly on agents or tools---checking outputs, state, memory, or reasoning. Other tests assert on the environment or non-agentic components within the system. We also observe a small number of tests without explicit assertions.

\begin{table}[t]
\centering
\caption{Distribution of testing patterns in agent-related tests.}
\label{tab:testing-patterns}
\tiny
\setlength{\tabcolsep}{3pt}
\resizebox{0.77\columnwidth}{!}{%
\begin{tabular}{p{0.2\columnwidth}p{0.4\columnwidth}rr}
\toprule
\textbf{Category} & \textbf{Pattern} & \textbf{Count} & \textbf{\%} \\
\midrule
\multirow{8}{=}{\textbf{Fixture Patterns}}
& $\bullet$ Mock agent/tool    & 886  & 34.45 \\
& \hspace{0.5em}$\circ$ \hspace{0.5em}Mock some component & 557  & 21.66 \\
& \hspace{0.5em}$\circ$ \hspace{0.5em}Mock entire agent & 573  & 22.28 \\
& $\bullet$ Mock environment   & 241  & 9.37 \\
& \hspace{0.5em}$\circ$ \hspace{0.5em}Mock client/server & 156  & 6.07 \\
& \hspace{0.5em}$\circ$ \hspace{0.5em}Mock configuration & 85   & 3.3 \\
& $\bullet$ Mock external deps. & 137  & 5.33 \\
& $\bullet$ Mock non-agentic system & 132  & 5.13 \\
\midrule

\multirow{4}{=}{\textbf{Data Patterns}}
& $\bullet$ Realistic synthetic data & 1390 & 54.04 \\
& $\bullet$ Dummy data          & 1026 & 39.89 \\
& $\bullet$ Mocked data        & 98   & 3.81 \\
& $\bullet$ Real data          & 39   & 1.52 \\
\midrule

\multirow{9}{=}{\textbf{Objective Patterns}}
& $\bullet$ Test app agents/tools & 1825 & 70.96 \\
& $\bullet$ Test synthetic agents/tools & 470  & 18.27 \\
& $\bullet$ Test interaction   & 344  & 13.37 \\
& \hspace{0.5em}$\circ$ \hspace{0.5em}Environment & 74   & 2.88 \\
& \hspace{0.5em}$\circ$ \hspace{0.5em}Other agents/tools & 169  & 6.57 \\
& \hspace{0.5em}$\circ$ \hspace{0.5em}Non-agentic (non-det.) & 105  & 4.08 \\
& \hspace{0.5em}$\circ$ \hspace{0.5em}Non-agentic (det.) & 12   & 0.47 \\
& $\bullet$ Test scenario      & 269  & 10.46 \\
& $\bullet$ E2E testing        & 258  & 10.03 \\
\midrule

\multirow{8}{=}{\textbf{Assertion Patterns}}
& $\bullet$ Assert on agent/tool & 2292 & 89.11 \\
& \hspace{0.5em}$\circ$ \hspace{0.5em}Output & 1437 & 55.87 \\
& \hspace{0.5em}$\circ$ \hspace{0.5em}State & 1317 & 51.21 \\
& \hspace{0.5em}$\circ$ \hspace{0.5em}Reasoning & 225  & 8.75 \\
& \hspace{0.5em}$\circ$ \hspace{0.5em}Memory & 68   & 2.64 \\
& $\bullet$ Assert on environment & 283  & 11 \\
& $\bullet$ Assert on non-agent & 77   & 2.99 \\
& $\bullet$ No assertion       & 33   & 1.28 \\
\bottomrule
\end{tabular}%
}
\caption*{\scriptsize Note: Tests can belong to multiple patterns; percentages may exceed 100\% within categories.}
\vspace{-5pt}
\end{table}

\vspace{-5pt}
\subsection{Key Observations and Findings}
\label{subsec:test-pattern-findings}

Table~\ref{tab:testing-patterns} summarizes the results on testing patterns. Overall, we observe a broad distribution across the various testing patterns. Among mocking-related patterns, the largest share involves mocked agents or tools (34.4\%), followed by environment configuration (9.4\%) and external dependencies (5.3\%). A smaller fraction (5.1\%) mocks non-agentic systems, such as file systems or databases. Next, we highlight several interesting findings from these patterns.

\vspace{-5pt}
\paragraph{Test Data Creation} As mentioned earlier, tests often use simple example-style inputs (e.g., \ttsmall{test1}, \ttsmall{test2}). For example, the following test verifies whether personal information is removed from agent output when a policy related to personally identifiable information (PII) is enabled. However, it relies on a single sample record containing a name, email, and phone number. Given the broader scope of PII, this test provides limited evidence that the agent can reliably handle different types of sensitive information.

\begin{pythoncode}
def test_agent_policy_without_feedback():
    agent = Agent(model="openai/gpt-4o-mini",agent_policy=PIIBlockPolicy,...)
    assert agent.agent_policy is not None, "agent_policy should be set" ...
    re==agent.do("Generate a sample customer record with name, email and phone")
    output4=output_buffer.getvalue()
    assert any(keyword in res.lower() for keyword in ["pii", "personal", "identifiable", "blocked", "remove", "anonymize"])
\end{pythoncode}

This points to a broader gap in coverage and adequacy. Traditional software testing offers well-defined coverage hierarchies~\cite{ammann2016testing, zhu1997coverage, weyuker1988adequacy}, but agentic systems lack comparable criteria. Existing notions do not capture higher-level abstractions such as agents, tools, memory, environment, and their interactions, which are rarely formalized or measured. As a result, many agent tests rely on a few illustrative examples rather than systematic coverage.

\vspace{-3pt}
\begin{findingbox}{Test Data Creation}
39.9\% of tests use simple \change{dummy data consisting of} example-like inputs, suggesting test data is often illustrative rather than realistic or diverse.
\end{findingbox}
\vspace{-9pt}

\vspace{-5pt}
\paragraph{Component Mocking} 
Mocking is a common strategy for isolating agentic behavior during testing. Overall, 34.4\% of tests mock an agent or tool in some form: 21.7\% mock selected components, while 22.3\% mock the entire agent. This suggests that mocking is a common strategy for isolating agentic behavior during testing, although the way it is applied varies across projects.

We further analyzed the types of mocking used in the tests. Most mocking instances rely on standard mechanisms such as \ttsmall{unittest.mock} or \ttsmall{@fixture}, while only a small fraction (13.4\%) employ custom test doubles tailored to specific testing objectives. For example, in any-agent~\cite{any-agent}, the test \ttsmall{test\_a2a\_tool\_multiturn} uses a mocked agent backed by a mocked LLM against a real agent-to-agent server. This hand-crafted test double replaces non-deterministic LLM behavior with deterministic scripted responses while preserving the surrounding framework execution.

\begin{pythoncode}
class MockConversationAgent(TinyAgent):
    """Mock agent implementation that simulates multi-turn conversation."""
    async def _load_agent(self) -> None: ...
    async def run_async(self, prompt: str, ...) -> AgentTrace:
        # Overrides the real LLM call entirely
        if self.turn_count == 0: ...  # returns scripted first-turn response
        elif self.turn_count == 1: ...  # returns scripted second-turn response
        elif self.turn_count == 2: ...  # returns scripted third-turn response
    def _create_mock_trace() -> AgentTrace:
        """Create a mock AgentTrace with minimal spans for testing.""" ....
async def test_a2a_tool_multiturn() -> None:
    """Test agents can maintain context across multiple interactions."""
    # Create a mock agent that simulates multi-turn conversation
    config = AgentConfig(...)
    agent = MockConversationAgent(config)
    server_handle = await agent.serve_async(...)
    # Three real HTTP calls to the real A2A server - LLM is the only thing mocked
    response_1 = await client.send_message(request_1, ...)   # turn 0
    response_2 = await client.send_message(request_2, ...)   # turn 1
    response_3 = await client.send_message(request_3, ...)   # turn 2
    assert agent.turn_count == 3  # mock was hit exactly 3 times
    result = UserInfo.model_validate_json(response_3)
    assert result.age == 30
\end{pythoncode}
\change{The prevalence of existing mocking frameworks partly reflects the modular design of agents. Agents, tools, LLM clients, and external services typically expose clear interaction boundaries, allowing developers to isolate dependencies through conventional patching, fixtures, and interface replacement. Custom test doubles are mainly needed for domain-specific, stateful, or multi-turn behavior that simple return-value or exception-based mocks cannot capture.
}




\vspace{-5pt}
\paragraph{Test Assertions}
Most assertions target observable properties such as state (51.1\%) and output (55.9\%), whereas only 8.7\% address reasoning-related behavior. This gap is notable because reasoning is central to many agentic systems yet is rarely tested directly. Even when it is considered, tests usually check only for the presence of reasoning-like behavior rather than its correctness or usefulness. For example, a test from Upsonic~\cite{upsonic} verifies that subagent activity appears in the logs, but not whether delegation was appropriate or contributed correctly to the final result.

\begin{pythoncode}
async def test_deepagent_with_subagents():
    researcher = Agent(goal="Conduct thorough research on topics",)
    writer = Agent(goal="Write clear and engaging content",)
    agent = DeepAgent(name="Test DeepAgent",subagents=[researcher, writer],...)...
    task = Task(description=("Use the researcher subagent ..."))...
    assert result is not None
    assert "researcher" in output.lower() or "task" in output.lower()
\end{pythoncode}

This suggests that reasoning is largely treated as an implicit mechanism rather than an explicit test target. Although properties related to reasoning and coordination can be checked via simulation frameworks (e.g., LangChain~\cite{langchaintrajectory}, LangFuse~\cite{langfuse}, Arize AI~\cite{arizeai}, Maxim AI~\cite{maximai}) that expose execution trajectories, intermediate decisions, tool calls, and environment transitions, such analyses are typically external to the main test suite and not well supported by standard testing frameworks such as PyTest and Unittest. Consequently, even when developers inspect trajectories, these checks are rarely integrated into routine testing workflows. This creates a gap between what can be evaluated and what is systematically validated. Bridging this gap requires testing frameworks that support assertions over trajectories, delegation behavior, and intermediate reasoning steps—an important direction for future research.


\vspace{-3pt}
\begin{findingbox}{Test Assertions}
Only 8.7\% of tests assert on reasoning-related behavior, and even these tests typically check only the presence of such behavior rather than its quality or correctness.
\end{findingbox}
\vspace{-9pt}

\vspace{-5pt}
\paragraph{Interactions with Non-agentic Components}
Such interactions appear in a small but meaningful share of tests. These tests matter because agent correctness often depends on the surrounding software stack, not just the agent itself. For example, one test checks not only agent output but also whether the agent writes to a file and invokes an external service via \ttsmall{uvx} and \ttsmall{MCPStdio}. 

\begin{pythoncode}
def test_load_and_run_agent(...) -> None:
    def write_file(text: str) -> None: ...
    tools = [write_file,MCPStdio(command="uvx",args=["mcp-server-time", "--local-timezone=America/New_York"],tools=["get_current_time",],),] ...
    assert (tmp_path / tmp_file).read_text() == str(datetime.now().year)
\end{pythoncode}

\begin{findingbox}{Interactions with non-agentic parts}
    Only 4.1\% of the tests validate interactions with non-deterministic non-agentic components and 2.4\% with deterministic ones.
\end{findingbox}
\vspace{-9pt}

\subsection{Future Research Directions}
\label{subsec:test-pattern-future-direction}

Based on our findings on structural gaps in how agent tests are designed, we outline research directions that address the deficiencies. As with the testing-level gaps discussed in \S\ref{subsec:test-level-future-direction}, many of the challenges emerging have antecedents in the broader SE literature, but must be adapted to agentic systems.

\vspace{-5pt}
\paragraph{Test Isolation for Agentic Systems}

Mocking is a common isolation strategy in agentic systems: 34.4\% of the tests in our dataset mock agents or tools. Yet without reusable abstractions, developers often rely on ad-hoc fixtures, patching, and manual stubbing. This challenge goes beyond tooling. Traditional test-double theory~\cite{mackinnon2000endo} assumes deterministic behavior and well-typed interfaces, but agentic systems often violate both. In particular, agent--tool interactions expose a dual interface: a machine-facing API for invocation and a natural-language description that shapes LLM reasoning. Mocking a tool therefore requires simulating not only its execution behavior but also the effect of its description on the agent. This duality is especially visible in protocols such as MCP~\cite{mcp}, and recent work shows that tool metadata itself can influence agent behavior~\cite{hasan2026model}.
A second challenge is that correctness in agentic systems is often distributional rather than point-valued~\cite{dobslaw2025challenges}. Instead of reproducing one fixed output, tests may need to preserve acceptable behavior across repeated executions. As a result, useful test doubles may need to preserve output distributions rather than return fixed responses. In practice, developers often avoid this complexity by testing deterministic components while under-testing the non-deterministic core~\cite{hasan2025empirical}. Future work should identify the seams, interfaces, and abstractions needed to make agentic systems more testable, and formalize the properties test doubles must preserve.

\vspace{-5pt}
\paragraph{Environment State and Side-Effect Validation}

Only a small percentage of tests in our dataset validate interactions with external components, suggesting that current practice focuses more on outputs than on the actions agents perform. This is a major gap, as agents increasingly execute multi-step workflows that modify files, invoke services, and update databases. In traditional SE, state-based and property-based testing support validation of such effects~\cite{chow1978testing, hughes2007quickcheck}, but these techniques appear to be rarely used in agentic systems. Some recent benchmarks move in this direction: $\tau$-bench~\cite{yao2024tau} evaluates resulting database state, while OSWorld~\cite{xie2024osworld} inspects execution artifacts such as files and broader system state.
These observations suggest that environment state should become a first-class target of agent testing. Doing so will require methods for comparing pre- and post-execution states, reasoning about partially completed tasks, and maintaining consistency in shared environments where multiple agents may act concurrently.

\vspace{-5pt}
\paragraph{Systematic Test Input and Environment Generation}

Test design in agentic systems remains limited in both input breadth and environment depth. Our findings show that 39.9\% of tests use simple example-like inputs, while only 10.5\% cover multi-step scenarios. This suggests that many tests exercise only a narrow part of the input space in relatively simple settings. In traditional testing, this problem has been addressed through techniques for systematically exploring inputs, including model-based testing~\cite{utting2012taxonomy} and coverage-guided fuzzing~\cite{bohme2016coverage}. Recent work has started to adapt these ideas to agentic systems: ToolFuzz~\cite{milev2025toolfuzz} and ChainFuzzer~\cite{wu2026chainfuzzer} apply fuzzing to tool runtimes and multi-tool workflows, while other approaches explore property-based testing~\cite{maaz2025agentic} and adversarial search~\cite{wang2025agentvigil}.
However, significant gaps remain. Future work should treat test design for agentic systems as a first-class concern and develop scenario-generation techniques that combine realistic inputs with rich, representative environments.

\vspace{-5pt}
\section{Agent Test Complexity}
\label{sec:rq_agent_test_complexity}


\begin{description}[leftmargin=1.0em, labelsep=0.5em, itemsep=0pt, topsep=2pt]
    \item[RQ3 (Test Complexity):] What is the structural complexity of agent-related tests?
\end{description}

In this RQ, we analyze the structural complexity of agent-related tests. We define a test complexity model and use static analysis to extract key properties of test methods to populate it ~\cite{tangent_artifact}. At the assertion level, we record assertion types and code. We also collect standard complexity metrics, including non-comment lines of code (NCLOC) and cyclomatic complexity, together with call-related metrics such as constructor and library calls. At the method level, we extract information on mock usage, identify fixture methods (setup and teardown), and analyze helper methods, including their assertions. For each test, we record its name, location, associated fixtures and helpers, and aggregate counts such as the number of fixtures, helpers, mocks, and assertions.
Our analysis~\cite{tangent_artifact} shows that agent-related tests are generally structurally simple. The median test contains 14 NCLOC and has cyclomatic complexity one; including fixtures and helpers raises these medians only to 20 NCLOC and three. Assertions are also lightweight, with a median of two per test. 
Call statements are similarly limited, with medians of one constructor call and four library calls per test. 

\textit{Research Directions.} Our recent study of Java tests~\cite{pan2026hamster} found that real-world tests often exercise multiple focal classes and methods and validate interactions across components. Such interaction-oriented tests appear less common in current agentic systems, highlighting the need for techniques that generate tests beyond isolated behaviors and systematically exercise interactions among components, environments, and non-agentic parts of the application. \change{Future work should also develop agent-specific complexity metrics, as NCLOC may not adequately capture the complexity of agent tests. More suitable measures could consider factors such as the number of tool calls, agent interactions, orchestration logic, state, and external dependencies.}

\section{Perspective of Industry Practitioners}
\label{sec:rq_industry}
\begin{table}[t]
\centering
\caption{Interview Questionnaire.}
\label{tab:survey_questions}
\tiny
\setlength{\tabcolsep}{2pt}

\begin{minipage}{0.81\columnwidth}
\centering
\begin{tabularx}{\linewidth}{|X|}

\multicolumn{1}{c}{\textbf{Section 1: Practitioner Background and Organizational Context}} \\

\hline
Q1. What is your current role in your organization? \\
Q2. What types of agents does your team build or use? \\
Q3. What technologies or frameworks are used to build these agents? \\
Q4. In what domains or workflows are these agents primarily applied? \\
Q5. What is the scale of deployment? \\

\hline
\multicolumn{1}{c}{\textbf{Section 2: Agent Architecture and System Design}} \\
\hline
Q6. How are your agents and agent applications typically structured? \\
Q7. What are the agent implementation patterns you have used in the past? \\

\hline
\multicolumn{1}{c}{\textbf{Section 3: Testing Strategies and Methodologies}} \\
\hline
Q8. What components of the agent system do you consider to be most critical to test? \\
Q9. What types of testing do you perform for agents? \\
Q10. What types of tests do your agents fail most often? \\
Q11. How do you test agent reasoning or decision-making? \\
Q12. How do you identify bugs in your agents? \\
Q13. Do you use automated testing pipelines for agents? If so, how are they implemented? \\
Q14. Do you test live agents or tools or use mocking heavily to test components in isolation? \\

\hline

\multicolumn{1}{c}{\textbf{Section 4: Test Data and Scenario Design}} \\
\hline

Q15. Do you use real user queries, synthetic prompts, or curated benchmarks? \\
Q16. How do you ensure your test cases cover all scenarios effectively? What metrics do you use to measure coverage? \\
Q17. Do you simulate multi-step tasks or long conversations during testing? \\

\hline
\multicolumn{1}{c}{\textbf{Section 5: Monitoring and Continuous Evaluation}} \\
\hline
Q18. How do you monitor agent behavior after deployment? \\
Q19. Do you perform continuous evaluation or regression testing when models change? \\

\hline
\multicolumn{1}{c}{\textbf{Section 6: Non-functional Requirements}} \\
\hline
Q20. Do you test agents for non-functional requirements (NFR)? If yes, which NFRs? \\
Q21. What processes are in place to handle failures or unexpected agent behavior? \\

\hline
\multicolumn{1}{c}{\textbf{Section 7: Challenges}} \\
\hline
Q23. What are the biggest challenges in testing AI agents compared to traditional software? \\
Q24. Which aspects of agent behavior are hardest to test? \\
Q25. What is your advice to build testing pipelines for agents? \\

\hline
\multicolumn{1}{c}{\textbf{Section 8: Future of Software Agents}} \\
\hline
Q26. How do you see agent testing practices evolving in the next few years? \\
Q27. What tools or frameworks do you wish existed for better agent testing? \\
Q28. What role do you think automated evaluation or AI-based testing will play? \\
Q29. When agents are generated by other agents, what could be the role of testing? \\

\hline
\end{tabularx}
\end{minipage}

\label{tb:questionnaire}
\vspace{-10pt}
\end{table}

\begin{description}[leftmargin=1.0em, labelsep=0.5em, itemsep=0pt, topsep=2pt]
    \item[RQ4 (Practitioner Perspective):] How do industry practitioners perceive testing practices and challenges for agent applications?
\end{description}

In the previous RQs, we examined tests from open-source projects, identifying common patterns and gaps. In this RQ, we shift focus to industry practitioners building production-ready agentic solutions and platforms to understand their perspectives on testing practices and challenges for agent applications.
We analyzed 10 interview responses from developers and researchers working on agentic technologies at IBM. 
Table~\ref{tab:survey_questions} summarizes the questionnaire.
\change{We first discuss the participant selection criteria and industry experience, followed by the key observations.}

\textit{\change{Selection Criteria and Industry Experience}} 
\change{
We recruited industry practitioners with recent experience developing or deploying agent-based products, representing roles including AI engineers and architects, managers, senior technical staff and researchers, and an executive leading AI and agent products. Participants averaged over 17 years of industry experience, ranging from 5 to 29 years.
}

\textit{Section 1: Background and adoption.}
The responses span diverse agent use cases, including code refactoring, patch generation, code migration, software engineering assistance, and IT operations support, indicating adoption across a broad range of engineering and enterprise workflows beyond experimental settings. They also reflect varying levels of organizational maturity: 70\% report internal tool use, 50\% production deployment, and 40\% pilot deployment.

\textit{Section 2: Architecture and system design.}
No single architecture or framework dominates, but several patterns emerge. All responses (100\%) describe tool-calling systems; 90\% report single-agent designs, 80\% planning- or reasoning-based approaches, and 50\% multi-agent systems. Workflow- or graph-based designs appear in 40\% of the responses, as do retrieval- and memory-augmented structures. At the framework level, MCP is the most common, followed by FastMCP, LangGraph, and LangChain. Overall, these results suggest that industry agent systems are typically composite systems built around tool use, orchestration, and coordination.

\textit{Section 3: Testing strategies and methodology.}
Participants report a broad mix of testing strategies. Unit and integration testing are each used by 70\% of participants, and end-to-end testing is also common (60\%), whereas module-level (10\%) and API-level (20\%) testing are less frequent. The reported test targets range from individual components, such as tool-calling modules, orchestrator agents, and authorization mechanisms, to the full system. 
Responses on reasoning-related testing and bug identification also show substantial variation. Some participants rely on curated benchmarks, while others emphasize manual trajectory review, detailed happy-path and edge-case specifications, live sampling, differential observability, and state modeling. One participant summarized this broader approach as \textit{``Benchmarks, differential observability, and state modeling.''} This suggests that reasoning evaluation and debugging often extend beyond benchmark execution. One participant also reported more than 90\% code coverage for their framework, but clarified that this mainly applies to code-level components rather than user scenarios, motivating the need to examine test data more closely.

\textit{Section 4: Test data and adequacy.}
Most responses (70\%) refer to curated benchmarks or benchmark-style environments, while 30\% report using real user queries or interactions and another 30\% synthetic data generation. Together, these responses suggest that industry practice often adopts a hybrid test-data strategy shaped by domain needs and privacy constraints. Test adequacy, however, appears to lack any widely accepted standard. As one participant put it, \textit{``No metrics at this point.''} Instead, participants rely on local practices such as edge-case coverage, domain-driven test design, trajectory sampling, query clustering, and acceptable failure thresholds. Overall, adequacy is assessed through project-specific judgment rather than formal coverage criteria.

\textit{Sections 5 and 6: Post-deployment monitoring, regression testing, and NFRs.}
Because agent behavior cannot be fully validated offline, practitioners rely heavily on post-deployment observation. Tracing and observability are central: 50\% explicitly mention OpenTelemetry-style tracing, 70\% refer to observability more broadly, and participants also cite logs, Langfuse, and LangSmith, suggesting reliance on a broader observability stack rather than a single tool. In addition, 50\% report continuous evaluation or regression testing during model updates, typically through reruns, before--after comparisons, downstream performance evaluation, and trace inspection. NFR testing also receives greater emphasis than in the open-source projects we analyzed: 80\% of responses mention concerns such as security, trust, performance, reliability, latency, scalability, resilience, and load handling, with some also describing stress and chaos testing. 

\begin{table}
\centering
\caption{\change{Nature of Findings.}}
\label{tb:nature_of_findings}
\footnotesize
\setlength{\tabcolsep}{3pt}
\begin{tabular}{lccc}
\hline
\textbf{Findings} & 
\textbf{LT} & 
\textbf{LTP} & 
\textbf{GO} \\
\hline
Testing Level & X &  &  \\
Agentic Framework vs. Testing Level &  &  & X \\
Non-Functional Testing & X & X &  \\
Test Data Creation & X &  &  \\
Test Assertions &  & X &  \\
Interaction with Non-Agentic Parts &  & X &  \\
\hline
\end{tabular}\\
* LT: Lack of Theory, LTP: Lack of Theory-to-Practice, and GO: Generic Observation
\vspace{-9pt}
\end{table}

\textit{Sections 7 and 8: Challenges and future directions.}
The interviews reveal several recurring challenges and future needs in testing agentic systems. First, participants emphasized non-determinism as a central difficulty, noting that inconsistent behavior across runs complicates debugging and validation. Second, they also highlighted the challenge of representing diverse user personas in test suite design. Third, adequacy emerged as a major gap, with no widely accepted standard: one participant stated, \textit{``No metrics at this point,''} and noted the lack of \textit{``automatic test case generation, synthetic data generation, and also test coverage tools.''} Fourth, participants stressed the need for stronger specifications, with one highlighting \textit{``state modeling and testing formal specifications over those states.''} 
Fifth, they recommended an observability-first approach. 
Sixth, current practice still relies heavily on manual and benchmark-driven evaluation. 
Seventh, participants called testing \textit{``critical''} as agentic systems mature toward realistic, high-stakes deployment. Finally, they viewed automated evaluation as essential yet immature;
as one participant put it, \textit{``Automation of testing is completely absent. Users start every time from scratch for every single project.''} Taken together, these responses suggest that agentic system testing is becoming more important, but still lacks in several dimensions.

\vspace{-10pt}
\section{Research Roadmap}
\label{sec:discussion}
\begin{figure}[t]
    \centering
    \includegraphics[width=.7\linewidth]{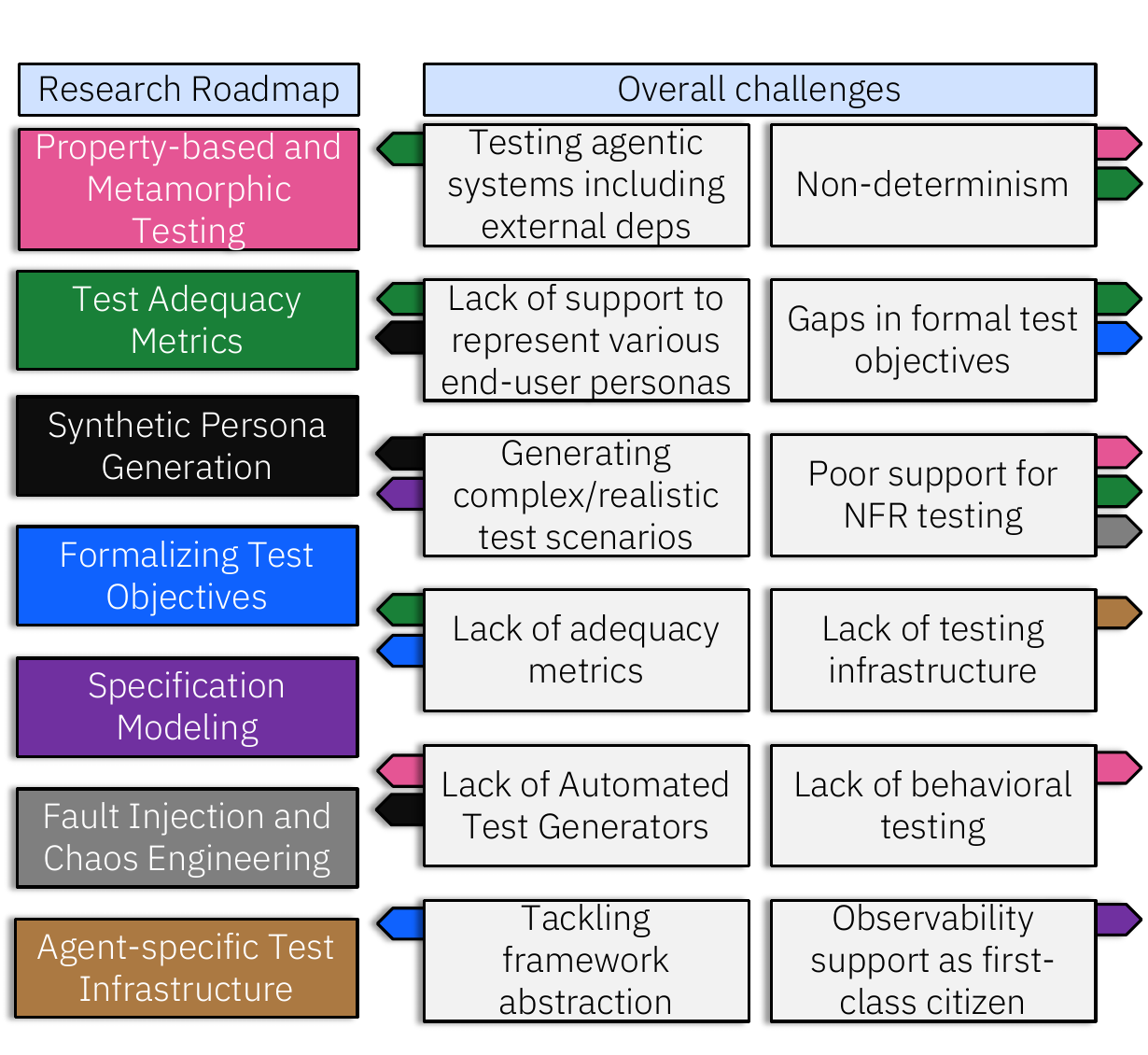}
    \vspace{-2pt}
    \caption{Research roadmap for addressing common challenges in agent application testing and validation.}
    \label{fig:roadmap}
    \vspace{-8pt}
\end{figure}

Building on our findings on the key challenges of testing agentic systems and on possible ways to address them, we propose a research roadmap for the SE community that highlights important open problems.
\change{We first summarize the key observations from our manual study in Table~\ref{tb:nature_of_findings}. We then incorporate insights from our practitioner interviews and present potential research directions, along with the high-level challenges they address, in Figure~\ref{fig:roadmap}.}

\change{
Most observations reflect either the absence of testing theory for agentic applications or the limited adaptation of existing theory. Although conventional testing levels are well established, agentic systems lack formal test objectives, clear testing boundaries, automated test generation, and adequacy metrics. Challenges such as NFR testing and test-data generation require both adapting theories for probabilistic software and developing new approaches to address agentic non-determinism and vast behavior spaces. Assertions and interactions with non-agentic components have stronger theoretical foundations, but still lack agent-specific infrastructure and expressive assertion mechanisms. Combining these findings with practitioner interviews, we identify several directions for future research.
}
One potential direction is \textit{Property-based and Metamorphic Testing}, motivated by the recurring challenge of non-determinism in agent applications, which makes testing harder than in traditional software. Related repeatability and oracle challenges have long been studied in SE, particularly for flaky tests~\cite{rahman2025understanding, dutta2020detecting} and probabilistic programs~\cite{dutta2018testing}. Prior work has addressed these challenges through \change{probabilistic model checking~\cite{jansen2016bounded, forejt2011automated}}, probabilistic symbolic execution~\cite{luckow2014symbolic,geldenhuys2012symbolic}, fuzzing~\cite{dutta2018testing}, differential testing~\cite{chen2016coverage}, property-based testing~\cite{hughes2007quickcheck}, and metamorphic testing~\cite{chen2018metamorphic}. \change{Collectively, these techniques provide precedents for supplementing exact-output oracles with invariants, quantitative constraints, and relations across executions or implementations, rather than requiring every execution to produce one prescribed output.}

Another direction is \textit{Test Adequacy Metrics}. Several possibilities include tool coverage, which measures whether the tools available to an agent and their execution paths are sufficiently exercised; persona coverage, which evaluates whether the test suite reflects the diversity of representative end users and their interactions; hybrid adequacy metrics spanning agentic and non-agentic components; and fault-based perspectives, in which fault coverage assesses whether a test suite reveals failures.
A further direction is \textit{Synthetic Persona Generation}. Because agentic systems are increasingly used by diverse end users, representative test data generation is especially challenging. For instance, a coding agent may serve both highly technical developers and business stakeholders, whose requirements and interaction patterns differ substantially, making representative test design difficult. Prior work on user profiling and modeling~\cite{purificato2024user} provides a useful foundation for this problem.
Another promising direction is \textit{Specification Modeling}. Our interviews suggest that stronger specifications could improve both testing and debugging, building on approaches widely used in earlier generations of software agents~\cite{spec1, spec2} and a few early works in LLM-based agentic systems~\cite{spec3, spec4}.
Other directions include \textit{Formalizing Test Objectives} and \textit{Fault Injection and Chaos Engineering}, which we discussed earlier.
Finally, \textit{Agent-specific Test Infrastructure} is needed. Existing platforms expose trajectories, tool calls, intermediate decisions, and environment transitions, but these are rarely treated as first-class test targets. Better infrastructure could make automated testing of agentic systems more practical and reusable.

\vspace{-12pt}
\section{Threats to Validity}
\label{sec:threats}

\textit{Internal Validity.}
A threat to internal validity is over-interpreting observed associations, which may reflect project characteristics such as domain, maturity, development process, or repository structure. To mitigate this, we present our findings as tendencies rather than causal effects and use the practitioner study to contextualize, rather than validate, the repository-based results. 
\change{We also examine library version as a potential confounding factor by analyzing its effect on testing levels. (\S\ref{sec:rq_agent_test_patterns}). Although our pattern analysis aggregates across versions, the taxonomy captures framework-agnostic practices such as mocking, unrealistic test data, and weak assertions. Studying how these practices evolve over time remains future work. We focus on encoded tests rather than observability-based validation. While both approaches are important, encoded tests explicitly capture execution steps, intermediate states, tool calls, and assertions, and are fundamental to systematic software testing.
}

\textit{External Validity.}
Our analysis focuses on Python GitHub repositories using popular agent-development frameworks and quality-oriented selection criteria, which may limit generalizability to other languages, proprietary systems, or frameworks.
We mitigate this by covering multiple widely used frameworks and incorporating practitioner perspectives. Another threat is that some tests may be AI-generated. 
Although we applied heuristics~\cite{prarena} and found no clear evidence in committed code, AI use may still have occurred during development; regardless, the resulting testing practices remain the responsibility of the application owners.

\textit{Construct Validity.}
Construct validity is limited by the lack of standardized testing ontologies for agentic systems, making some categories approximate. Our extraction rules, LLM-assisted filtering, and manually derived taxonomy may miss edge cases or introduce misclassifications. We mitigate these threats through multi-labeler review, discussion-based resolution, Cohen’s kappa to assess agreement, and a practitioner study that complements and contextualizes the repository evidence.

\vspace{-8pt}
\section{Related Work}
\label{sec:related_works}

\change{Empirical research on testing agentic systems remains limited. The closest study, by \citeauthor{hasan2025empirical}~\cite{hasan2025empirical}, mines tests from 39 open-source agent frameworks and 439 applications and manually analyzes 759 unit test functions. It identifies ten testing patterns and maps tested subjects to canonical architectural components. Our study differs in focus, granularity, and method. Rather than assigning each test a primary subject and largely traditional testing patterns, we examine every agentic component involved and how it is configured, invoked, and verified toward the test objective. We manually analyze 2,572 tests across testing levels and types, objectives, data, mocking and isolation strategies, and structural complexity, producing an agent-specific catalog of 23 patterns spanning fixtures, data, objectives, and assertions. We further interview ten industry practitioners to investigate adoption and practice and derive a research roadmap grounded in classical testing theory.} Another closely related study by \citeauthor{kohl2025automated}~\cite{kohl2025automated} report a case study on agent testing. 
Their findings, however, are limited to a single case centered on the Amazon Bedrock Converse API, and the authors themselves call for larger-scale studies of agent-related tests. Our work directly addresses this need with a large-scale empirical analysis of such tests.
More broadly, empirical studies have examined testing practices for LLM-powered software systems~\cite{magalhaes2025testing}, but not specifically for agentic applications or agent-specific constructs such as tools, workflows, and multi-step interactions.

A larger body of work focuses on benchmarking agent capabilities. Benchmarks such as AgentBench~\cite{liu2023agentbench}, WebArena~\cite{zhou2023webarena}, SWE-bench and its variants~\cite{jimenez2023swebench, aleithan2024swebench_plus, deng2025swebench_pro}, and MultiAgentBench~\cite{zhu-etal-2025-multiagentbench} evaluate task completion under benchmark-defined metrics, but they do not show how developers encode expected behavior, robustness checks, and regressions in real tests. Related work on debugging and failure analysis studies failure taxonomies, fault attribution, bugs in orchestration frameworks, and interactive support for inspecting agent executions~\cite{cemri2026multi, zhang2025agent, xue2025characterization, epperson2025interactive, rorseth2025ladybug}. These works focus on diagnosing or repairing failures after execution.
\vspace{-7pt}
\section{Summary and Future Work}
\label{sec:conclusion}

We presented a comprehensive empirical study of testing practices for LLM-based agent applications, combining large-scale analysis of open-source projects with insights from industry practitioners. Our findings show that current testing is largely unit-focused, narrowly scoped, and provide limited support for non-functional validation and complex agent behaviors. We also identified common testing patterns and structural simplicity in test design. These results highlight significant gaps in current practices and motivate the need for more systematic and robust testing methodologies. To that end, we outlined research directions aimed at advancing the foundations, techniques, and tools for testing agent systems. Our publicly available dataset~\cite{tangent_artifact} supports future research in this area. In future work, we plan to pursue several of the identified research directions, including formalizing test objectives, defining test adequacy criteria, and developing test generation techniques for agentic applications.

\noindent\textbf{Data Availability.} The entire Tangent dataset including agentic labeling, static analysis, and the final labels are available at ~\cite{tangent_artifact}.

\bibliographystyle{ACM-Reference-Format}
\bibliography{references}

\end{document}